
\newif\ifproofmode			
\proofmodefalse				

\newif\ifforwardreference		
\forwardreferencetrue			

\newif\ifeqchapternumbers		
\eqchapternumbersfalse			

\newif\ifsectionnumbers			
\sectionnumberstrue			

\newif\ifeqsectionnumbers		
\eqsectionnumbersfalse			

\newif\ifchaptersectionnumbers     	
\chaptersectionnumberstrue		

\newif\ifcontinuoussectionnumbers	
\continuoussectionnumbersfalse	

\newif\ifcontinuousnumbers		
\continuousnumbersfalse 		

\newif\iffigurechapternumbers		
\figurechapternumbersfalse		

\newif\ifcontinuousfigurenumbers	
\continuousfigurenumbersfalse		

\newif\ifcontinuousreferencenumbers     
\continuousreferencenumberstrue         

\newif\ifparenequations			
\parenequationstrue			

\newif\ifstillreading			

\font\eqsixrm=cmr6			
\def\marginstyle{\eqsixrm}		

\newtoks\chapletter			
\newcount\chapno			
\newcount\sectno			
\newcount\eqlabelno			
\newcount\figureno			
\newcount\referenceno			
\newcount\minutes			
\newcount\hours				

\newread\labelfile			
\newwrite\labelfileout			
\newwrite\allcrossfile			

\chapno=0
\sectno=0
\eqlabelno=0
\figureno=0


%
\def\initialeqmacro{
    \ifproofmode
        \headline{\tenrm \today\ --\ \timeofday\hfill
                         \jobname\ --- draft\hfill\folio}
        \hoffset=-1cm
        \immediate\openout\allcrossfile=zallcrossreferfile
    \fi
    \ifforwardreference
        \openin\labelfile=zlabelfile
        \ifeof\labelfile
        \else
            \stillreadingtrue
            \loop
                \read\labelfile to \nextline
                \ifeof\labelfile
                    \stillreadingfalse
                \else
                    \nextline
                \fi
                \ifstillreading
            \repeat
        \fi
        \immediate\openout\labelfileout=zlabelfile
    \fi}


{\catcode`\^^I=9
\catcode`\ =9
\catcode`\^^M=9
\endlinechar=-1
\globaldefs=1


%
\def\chapfolio{			
    \ifnum \chapno>0 \relax
        \the\chapno
    \else
        \the\chapletter
    \fi}

%
\def\bumpchapno{
    \ifnum \chapno>-1 \relax
        \global \advance \chapno by 1
    \else
        \global \advance \chapno by -1 \setletter\chapno
    \fi
    \ifcontinuousnumbers
    \else
        \global\eqlabelno=0
    \fi
    \ifcontinuousfigurenumbers
    \else
        \global\figureno=0
    \fi
    \ifcontinuousreferencenumbers
    \else
        \global\referenceno=0
    \fi
    \sectno=0}

\def\bumpsectno{
    \global\advance\sectno by 1 \relax
    \ifeqsectionnumbers
        \ifcontinuoussectionnumbers
        \else
            \global\eqlabelno=0
        \fi
    \fi}

%
\def\setletter#1{\ifcase-#1 {}  \or\global\chapletter={A}
  \or\global\chapletter={B} \or\global\chapletter={C} \or\global\chapletter={D}
  \or\global\chapletter={E} \or\global\chapletter={F} \or\global\chapletter={G}
  \or\global\chapletter={H} \or\global\chapletter={I} \or\global\chapletter={J}
  \or\global\chapletter={K} \or\global\chapletter={L} \or\global\chapletter={M}
  \or\global\chapletter={N} \or\global\chapletter={O} \or\global\chapletter={P}
  \or\global\chapletter={Q} \or\global\chapletter={R} \or\global\chapletter={S}
  \or\global\chapletter={T} \or\global\chapletter={U} \or\global\chapletter={V}
  \or\global\chapletter={W} \or\global\chapletter={X} \or\global\chapletter={Y}
  \or\global\chapletter={Z}\fi}

%
\def\tempsetletter#1{\ifcase-#1 {}\or{} \or\chapletter={A} \or\chapletter={B}
 \or\chapletter={C} \or\chapletter={D} \or\chapletter={E}
  \or\chapletter={F} \or\chapletter={G} \or\chapletter={H}
   \or\chapletter={I} \or\chapletter={J} \or\chapletter={K}
    \or\chapletter={L} \or\chapletter={M} \or\chapletter={N}
     \or\chapletter={O} \or\chapletter={P} \or\chapletter={Q}
      \or\chapletter={R} \or\chapletter={S} \or\chapletter={T}
       \or\chapletter={U} \or\chapletter={V} \or\chapletter={W}
        \or\chapletter={X} \or\chapletter={Y} \or\chapletter={Z}\fi}

%
\def\chapshow#1{
    \ifnum #1>0 \relax
        #1
    \else
        {\tempsetletter{\number#1}\the\chapletter}
    \fi}

%
\def\today{\number\day\space \ifcase\month\or Jan\or Feb\or
        Mar\or Apr\or May\or Jun\or Jul\or Aug\or Sep\or
        Oct\or Nov\or Dec\fi, \space\number\year}

\def\timeofday{\minutes=\time    \hours=\time
        \divide \hours by 60
        \multiply \hours by 60
        \advance \minutes by -\hours
        \divide \hours by 60
        \ifnum\the\minutes>9 \relax
     		\the\hours:\the\minutes
 	\else
  		\the\hours:0\the\minutes
	\fi}


%
%
%
%

\def\chaplabel#1{
    \ifforwardreference                             
        \write\labelfileout{                        
        \noexpand\expandafter\noexpand\def          
        \noexpand\csname CHAPLABEL#1\endcsname{\the\chapno}}
    \fi
    \global\expandafter\edef\csname CHAPLABEL#1\endcsname
    {\the\chapno}
    \ifproofmode
        \rlap{\hbox{\marginstyle #1\ }}
    \fi}

%
\def\sectnum{
    \bumpsectno
        \ifchaptersectionnumbers
            \chapfolio.
        \fi
    \the\sectno}

\def\sectlabel#1{
    \bumpsectno
    \ifforwardreference
        \immediate\write\labelfileout{
        \noexpand\expandafter\noexpand\def
        \noexpand\csname SECTLABEL#1\endcsname{\the\chapno.\the\sectno?!}}
    \fi
    \global\expandafter\edef\csname SECTLABEL#1\endcsname
    {\the\chapno.\the\sectno?!}	 			
    \ifproofmode
        \llap{\hbox{\marginstyle #1\ }}
    \fi
    \ifchaptersectionnumbers
        \chapfolio.
    \fi
    \the\sectno}

\def\sectref#1{                                  
    \ifundefined{SECTLABEL#1}                     
        ++                                        
        \ifproofmode
            \ifforwardreference
            \else
                \write16{ ***Undefined\space Section\space Reference #1*** }
            \fi
        \else
            \write16{ ***Undefined\space Section\space Reference #1*** }
        \fi
    \else
        \edef\LABxx{\getlabel{SECTLABEL#1}}
	\ifchaptersectionnumbers
            \def\LAByy{\expandafter\stripchap\LABxx}
	    \chapshow\LAByy.
	\fi
	\expandafter\stripsect\LABxx
    \fi
    \ifproofmode
        \write\allcrossfile{Section\space #1}
    \fi}

%
%
\def\eqnum{                                    
    \global\advance\eqlabelno by 1              
    \eqno(
    \ifeqchapternumbers
        \chapfolio.
    \fi
    \ifeqsectionnumbers
        \the\sectno.
    \fi
    \the\eqlabelno)}

\def\eqlabel#1{                                
    \global\advance\eqlabelno by 1              
    \ifforwardreference                     
        \immediate\write\labelfileout{\noexpand\expandafter\noexpand\def
        \noexpand\csname EQLABEL#1\endcsname
        {\the\chapno.\the\sectno?\the\eqlabelno!}}
    \fi
    \global\expandafter\edef\csname EQLABEL#1\endcsname
    {\the\chapno.\the\sectno?\the\eqlabelno!}
    \eqno(
    \ifeqchapternumbers
        \chapfolio.
    \fi
    \ifeqsectionnumbers
        \the\sectno.
    \fi
    \the\eqlabelno)
    \ifproofmode
        \rlap{\hbox{\marginstyle #1}}		
    \fi}

\def\eqalignnum{                               
    \global\advance\eqlabelno by 1              
    &(\ifeqchapternumbers
        \chapfolio.
    \fi
    \ifeqsectionnumbers
        \the\sectno.
    \fi
    \the\eqlabelno)}

\def\eqalignlabel#1{                   	
    \global\advance\eqlabelno by 1 	        
    \ifforwardreference                     
        \immediate\write\labelfileout{\noexpand\expandafter\noexpand\def
        \noexpand\csname EQLABEL#1\endcsname
        {\the\chapno.\the\sectno?\the\eqlabelno!}}
    \fi
    \global\expandafter\edef\csname EQLABEL#1\endcsname
    {\the\chapno.\the\sectno?\the\eqlabelno!}
    &(\ifeqchapternumbers
        \chapfolio.
    \fi
    \ifeqsectionnumbers
        \the\sectno.
    \fi
    \the\eqlabelno)
    \ifproofmode
        \rlap{\hbox{\marginstyle #1}}			
    \fi}

\def\dnum{                                     
    \global\advance\eqlabelno by 1              
    \llap{(	 				
    \ifeqchapternumbers
        \chapfolio.
    \fi
    \ifeqsectionnumbers
        \the\sectno.
    \fi
    \the\eqlabelno)}}

\def\dlabel#1{                                 
    \global\advance\eqlabelno by 1              
    \ifforwardreference                         
        \immediate\write\labelfileout{\noexpand\expandafter\noexpand\def
        \noexpand\csname EQLABEL#1\endcsname
        {\the\chapno.\the\sectno?\the\eqlabelno!}}
    \fi
    \global\expandafter\edef\csname EQLABEL#1\endcsname
    {\the\chapno.\the\sectno?\the\eqlabelno!}
    \llap{(
    \ifeqchapternumbers
        \chapfolio.
    \fi
    \ifeqsectionnumbers
        \the\sectno.
    \fi
    \the\eqlabelno)}
    \ifproofmode
        \rlap{\hbox{\marginstyle #1}}		
    \fi}

\def\eqref#1{\ifparenequations(\fi
    \ifundefined{EQLABEL#1}***
        \ifproofmode
            \ifforwardreference
            \else
                \write16{ ***Undefined\space Equation\space Reference #1*** }
            \fi
        \else
            \write16{ ***Undefined\space Equation\space Reference #1*** }
        \fi
    \else
        \edef\LABxx{\getlabel{EQLABEL#1}}
	\def\LAByy{\expandafter\stripsect\LABxx}
        \def\LABzz{\expandafter\stripchap\LABxx}
        \ifeqchapternumbers
            \chapshow{\LABzz}.
        \else
            \ifnum \number\LABzz=\chapno \relax
            \else
                \chapshow{\LABzz}.
            \fi
        \fi
	\ifeqsectionnumbers
	    \LAByy.
	\fi
        \expandafter\stripeq\LABxx
    \fi
    \ifparenequations)\fi
    \ifproofmode
        \write\allcrossfile{Equation\space #1}
    \fi}

%
\def\fignum{                                   
    \global\advance\figureno by 1\relax         
    \iffigurechapternumbers
        \chapfolio.
    \fi
    \the\figureno}

\def\figlabel#1{				
    \global\advance\figureno by 1\relax 	
    \ifforwardreference				
        \immediate\write\labelfileout{\noexpand\expandafter\noexpand\def
        \noexpand\csname FIGLABEL#1\endcsname
        {\the\chapno.\the\sectno?\the\figureno!}}
    \fi
    \global\expandafter\edef\csname FIGLABEL#1\endcsname
    {\the\chapno.\the\sectno?\the\figureno!}
    \iffigurechapternumbers
        \chapfolio.
    \fi
    \ifproofmode
        \llap{\hbox{\marginstyle #1\ }}\relax
    \fi
    \the\figureno}

\def\figref#1{					
    \ifundefined				
        {FIGLABEL#1}!!!!			
        \ifproofmode
            \ifforwardreference
            \else
                \write16{ ***Undefined\space Figure\space Reference #1*** }
            \fi
        \else
            \write16{ ***Undefined\space Figure\space Reference #1*** }
        \fi
    \else
        \edef\LABxx{\getlabel{FIGLABEL#1}}
        \def\LABzz{\expandafter\stripchap\LABxx}
        \iffigurechapternumbers
            \chapshow{\LABzz}.\expandafter\stripeq\LABxx
        \else \ifnum\number\LABzz=\chapno \relax
                \expandafter\stripeq\LABxx
            \else
                \chapshow{\LABzz}.\expandafter\stripeq\LABxx
            \fi
        \fi
        \ifproofmode
            \write\allcrossfile{Figure\space #1}
        \fi
    \fi}

%
%
\def\pagelabel#1{
    \ifforwardreference
        \write\labelfileout{
        \noexpand\expandafter\noexpand\def
        \noexpand\csname PGLABEL#1\noexpand\endcsname{\the\pageno}}
    \fi
    \global\expandafter\edef\csname PGLABEL#1\endcsname{\the\pageno}}

\def\pageref#1{
    \ifundefined
        {PGLABEL#1}***
        \ifproofmode
        \else
            \write16{ ***Undefined\space Page\space Reference #1*** }
        \fi
    \else
        \csname PGLABEL#1\endcsname
    \fi
    \ifproofmode
        \write\allcrossfile{Page\space #1}
    \fi}

%
\def\refnum{                                      
    \global\advance\referenceno by 1\relax         
    \the\referenceno}	                           

\def\internalreflabel#1{			
    \global\advance\referenceno by 1\relax 	
    \ifforwardreference				
        \immediate\write\labelfileout{\noexpand\expandafter\noexpand\def
        \noexpand\csname REFLABEL#1\endcsname
        {\the\chapno.\the\sectno?\the\referenceno!}}
    \fi
    \global\expandafter\edef\csname REFLABEL#1\endcsname
    {\the\chapno.\the\sectno?\the\figureno!}
    \ifproofmode
        \llap{\hbox{\marginstyle #1\hskip.5cm}}\relax
    \fi
    \the\referenceno}

\def\internalrefref#1{				
    \ifundefined				
        {REFLABEL#1}!!!!			
        \ifproofmode
            \ifforwardreference
            \else
                \write16{ ***Undefined\space Footnote\space Reference #1*** }
            \fi
        \else
            \write16{ ***Undefined\space Footnote\space Reference #1*** }
        \fi
    \else
        \edef\LABxx{\getlabel{REFLABEL#1}}
        \def\LABzz{\expandafter\stripchap\LABxx}
        \expandafter\stripeq\LABxx
        \ifproofmode
            \write\allcrossfile{Reference\space #1}
        \fi
    \fi}

%
\def\reflabel#1{\item{\internalreflabel{#1}.}}

%
\def\refref#1{\internalrefref{#1}}

\def\eq{\ifhmode Eq.~\else Equation~\fi}		
\def\eqs{\ifhmode Eqs.~\else Equations~\fi}

%
%
%
%

%
\def\getlabel#1{\csname#1\endcsname}
\def\ifundefined#1{\expandafter\ifx\csname#1\endcsname\relax}
\def\stripchap#1.#2?#3!{#1}			
\def\stripsect#1.#2?#3!{#2}			%
\def\stripeq#1.#2?#3!{#3}			
}  

\overfullrule = 0pt
\magnification = 1200
\baselineskip 21pt
\forwardreferencetrue
\initialeqmacro
\def\sh{\mathop{\rm sh}\nolimits}
\def\ch{\mathop{\rm ch}\nolimits}
\def\coth{\mathop{\rm coth}\nolimits}

\line{\hfill LPTHE-PAR 94-04, UMTG-174}

\vskip 0.2 in

\centerline{\bf Scalar Kinks}

\bigskip

\medskip

\centerline{H. J. de Vega\footnote{$\dagger$}{Laboratoire de Physique
Th\'eorique et Hautes Energies, Tour 16 - 1er. \'etage, Universit\'e
Paris VI, 4, place Jussieu, 75252 Paris Cedex 05, FRANCE},
Luca Mezincescu${}^*$, and
Rafael I. Nepomechie\footnote*{Department of Physics, University of Miami,
Coral Gables, FL 33124, USA}}

\vskip 0.2 in

\bigskip

\centerline{\bf Abstract}

\vskip 0.2 in

We determine the excitations and $S$ matrix of an integrable isotropic
antiferromagnetic quantum spin chain of alternating spin 1/2 and spin 1.
There are two types of gapless one-particle excitations: the usual
spin 1/2 (``spinor'') kink, and a new spin 0 (``scalar'') kink.
Remarkably, the scalar-spinor scattering is nontrivial, yet the
spinor-spinor scattering is the same as if the scalar kinks were absent.
Moreover, there is no scalar-scalar scattering.

\bigskip

\bigskip

\vfill\eject

\noindent
{\bf 1. Introduction}

\bigskip

One-dimensional integrable quantum spin chains are among the few many-body
quantum systems for which some physical quantities can be calculated exactly.
The prototypical example is the antiferromagnetic $su(2)$-invariant
spin 1/2 Heisenberg chain${}^{\refref{bethe}}$.
At zero temperature ($T$) and zero magnetic field ($H$), the ground state is
described by an infinite filled Fermi sea, and the excitations consist of an
even number of spin 1/2 quasiparticles, called kinks. The dispersion
relation and $S$ matrix for the kinks have been
calculated${}^{\refref{faddeev/takhtajan}, \refref{korepin}}$. Moreover, the
thermodynamic properties of the model for low $T$ and small $H$ have been
obtained${}^{\refref{takahashi1}-\refref{takahashi2}}$.
It was conjectured in Ref. \refref{takahashi2}, and subsequently established,
that the specific heat ($C_H$) has the following property:
$$ \lim_{T \rightarrow 0} \lim_{H \rightarrow 0} {C_H\over T} =
\lim_{H\rightarrow 0} \lim_{T \rightarrow 0} {C_H\over T}
\,. \eqlabel{commute} $$
Indeed, the LHS can be evaluated by the method of
Filyov, {\it et al.}${}^{\refref{filyov}}$ (see, e.g., Ref. \refref{babujian}),
while the RHS can be evaluated by the method of Johnson and
McCoy${}^{\refref{johnson/mccoy}}$.

This model is critical. It can be regarded as a lattice field theory, which
in the continuum limit is a conformal field theory${}^{\refref{bpz}}$.
The central charge ($c$) is proportional${}^{\refref{critical}}$
to $\lim_{T \rightarrow 0} C_H/T$, and thus has the value $c=1$. The
conformal field theory in question has been identified${}^{\refref{critical}}$
as the level-one $SU(2)$ WZW model${}^{\refref{wzw}}$. We emphasize that,
in contrast to the bootstrap approach${}^{\refref{zamo}}$ for integrable
quantum field theories, here the $S$ matrix of the (gapless) kinks is
obtained by starting from the ``microscopic'' spin-chain Hamiltonian and the
corresponding Bethe Ansatz equations.

A major triumph of the quantum inverse scattering method is the construction
and solution of integrable isotropic higher-spin generalizations of the
Heisenberg chain, to which we shall refer as spin $s$
chains${}^{\refref{zamolodchikov/fateev}-\refref{takhtajan}}$.
For $s>1/2$, $C_H$ does {\it not} have the property \eqref{commute}.
The LHS (evaluated${}^{\refref{babujian}}$ by the method of
Filyov, {\it et al.}) gives $c= 3s/(s+1)$,
while the RHS (evaluated${}^{\refref{unpub}}$ by the method of Johnson-McCoy)
gives $c=1$. \footnote*{We remark that the central charge can also be
determined from finite-size corrections of the ground-state energy.
As shown in Ref. \refref{alcaraz/martins},
an analytic computation of finite-size corrections for the spin $s$ chain
which makes use of the string hypothesis gives $c=1$, which is
consistent with the results for $T=0 \,, H=0^+$; while the corresponding
numerical calculation (which does not make use of the string hypothesis)
gives $c= 3s/(s+1)$, which is consistent with the
results for $H=0 \,, T=0^+$. For $s=1$, analytic calculations not using the
string hypothesis also give${}^{\refref{klumper}}$ $c=3/2$.
It is therefore tempting to conjecture that the use of the string hypothesis
in the analytic finite-size scaling computation has the implicit
assumption $H=0^+$.}

Evidently, thermodynamic properties of $s > 1/2$ chains at the point
$T=H=0$ depend on how the point is approached in the $(T\,, H)$ plane,
e.g., $H=0 \,, T=0^+$ versus $T=0 \,, H=0^+$. Moreover, there is a discrepancy
between the results of Takhtajan${}^{\refref{takhtajan}}$ (see also Ref.
\refref{andrei/destri}) and Reshetikhin${}^{\refref{reshetikhin}}$ for
the two-body $S$ matrix. These facts strongly suggest that
there are (at least) two continuous field theories in the
$(T\,, H) = (0 \,, 0)$ limit of the spin $s$ isotropic chain.
A field theory with Takhtajan's $S$ matrix corresponds to the limit
$T=0 \,, H=0^+$ \footnote\dag{Takhtajan performs a quantum mechanical
(as opposed to thermodynamical) calculation, and thus, he clearly works at
$T=0$. The magnetic field $H$ does not appear explicitly in his calculation.
However, his calculation makes essential use of the string hypothesis, which
is not strictly correct for $T = H = 0$. (See, e.g., Ref.
\refref{destri/lowenstein}.) Since the string hypothesis is believed to be true
for max $(T \,, H) > 0$, his calculation is valid presumably only for
$H=0^+$.}; while a field theory with Reshetikhin's $S$ matrix
corresponds to the limit $H=0 \,, T=0^+$. We hope to return to this
matter in the future.

The singular behavior at $T=H=0$ is presumably related to the RSOS structure of
the space of states, which
appears${}^{\refref{faddeev/reshetikhin},\refref{reshetikhin}}$ for $s > 1/2$
in both of the limits discussed above. Evidence
for such a structure is readily seen by counting
states. Indeed, since the kinks have spin 1/2, naively one expects that
the number of $n$-kink states is $2^n$. However, the Bethe Ansatz implies
that the actual number of such states is larger. For instance, for $s=1$, there
are $2^{(3n-2)/2}$ such states.

A new type of integrable quantum spin chain has recently been
constructed${}^{\refref{devega/woynarovich}}$,
involving spins of different types. In Ref. \refref{we} we have investigated
thermodynamic properties for the particular case of an $su(2)$-invariant chain
of alternating spin 1/2 and spin 1. More general cases have also been
studied${}^{\refref{aladim/martins1}-\refref{pfeffer}}$.
In this Letter, we perform a direct Bethe-Ansatz calculation of the $S$ matrix
for the excitations of the $su(2)$-invariant alternating spin 1/2 and spin 1
chain, at the conformal point.\footnote*{In Ref. \refref{we},
we consider a two-parameter $( \bar c \,, \tilde c )$ family of Hamiltonians.
For simplicity, we restrict our attention here to the case
$\bar c = \tilde c > 0$, for which there is a single speed of sound.} At first
thought, one expects that -- as for the $s > 1/2$ chains -- there may be
more than one possible field theory in the $(T\,, H) = (0 \,, 0)$ limit;
and hence, there may be more than one possible result for the $S$ matrix.
However, for this chain,
$C_H$ {\it does} satisfy the property \eqref{commute}. Indeed, the limits
on the RHS and LHS have been computed in Refs. \refref{we} and
\refref{aladim/martins1}, respectively, and give the same value of the
central charge ($c=2$). This suggests that the $S$ matrix for the
particular case of alternating spin 1/2 and spin 1 is unique.

Our results are surprising. In looking for 2-particle excitations, we find,
in addition to the expected singlet and triplet states, an additional
singlet state. We interpret this to mean that there are two types of
one particle excitations: the usual spin 1/2 (``spinor'') kink, and a
new spin 0 (``scalar'') kink. The kinks obey certain super-selection rules.
The scattering between spinor kinks is the same as for the
Heisenberg chain, and there is no scattering between scalar kinks.
Nevertheless, there is nontrivial scattering between the spinor kinks
and the scalar kinks.

\bigskip

\noindent
{\bf 2. Ground state and excitations}

\bigskip

We consider a system with a strictly alternating arrangement of $2N$ spins,
with spins 1/2 at even sites and spins 1 at odd sites. That is, there are
$N$ spins
${1\over 2}\vec \sigma_2 \,, {1\over 2}\vec \sigma_4 \,, \cdots \,,
{1\over 2}\vec \sigma_{2N}$
of spin 1/2 and $N$ spins $\vec s_1 \,, \vec s_3 \,,$
$\cdots \,, \vec s_{2N-1}$  of spin 1.
The $su(2)$-invariant Hamiltonian ${\cal H}$ is
given by${}^{\refref{devega/woynarovich},\refref{we}}$
$$\eqalignno{
{\cal H} =  -{1 \over 18} \sum_{n=1}^N \Big\{ &
\left( 2 \vec \sigma_{2n} \cdot \vec s_{2n+1} + 1 \right)
\left( 2 \vec \sigma_{2n+2} \cdot \vec s_{2n+1} + 3 \right) \cr
& + \left( 2 \vec \sigma_{2n} \cdot \vec s_{2n-1} + 1 \right)
\left[ \left( 1 +  \vec s_{2n-1} \cdot \vec s_{2n+1}\right)
\left( 2 \vec \sigma_{2n} \cdot \vec s_{2n+1} + 1 \right) + 2 \right]
\Big\} \,. \eqalignlabel{hamiltonian} \cr} $$
Note that the Hamiltonian contains both nearest
and next-to-nearest neighbor interactions. We assume periodic boundary
conditions: $\vec \sigma_{2n} \equiv \vec \sigma_{2n + 2N}$ and
$\vec s_{2n+1} \equiv \vec s_{2n +1 + 2N}$.

The corresponding energy, momentum, and spin eigenvalues are given
by${}^{\refref{devega/woynarovich}}$
$$
E = {7\over 12}N - \sum_{j=1}^M \left(
{1\over 2}{1\over \lambda_j^2 + {1\over 4}}
+ {1\over \lambda_j^2 + 1} \right) \,, \eqlabel{energy}  $$
$$
P = {1\over 2i} \sum_{j=1}^M \log
\left( {\lambda_j + {i\over 2} \over \lambda_j - {i\over 2}}
       {\lambda_j + i \over \lambda_j - i} \right) \,,
\eqlabel{momentum} $$
$$S^z = {3\over 2}N - M \,, \eqlabel{spin} $$
where the variables $\lambda_j$ satisfy the Bethe Ansatz (BA) equations
$$
\left( {\lambda_j + {i\over 2} \over \lambda_j - {i\over 2}}
       {\lambda_j + i \over \lambda_j - i} \right)^N
= - \prod_{k=1}^M {\lambda_j - \lambda_k + i \over
\lambda_j - \lambda_k - i } \,, \qquad j = 1, \cdots , M \,. \eqlabel{BA} $$
The momentum operator
is defined as one-half the log of the two-site shift operator, and
hence the factor $1/2$ in \eq\eqref{momentum}. The Bethe Ansatz states are
highest weight vectors of $su(2)$ (see, e.g., Ref. \refref{faddeev/takhtajan}),
and thus have spin quantum numbers $S = S^z \ge 0$.

We assume the string hypothesis, which states that the solutions of the BA
equations \eqref{BA} for $N \rightarrow \infty$
are collections of $M_n$ strings of length $n$ of the form
(for $M_n > 0$)
$$\lambda_\alpha^{(n,j)} = \lambda_\alpha^n + i \left( {n+1\over 2} - j \right)
\,, \eqlabel{string} $$
where $j = 1, \cdots , n$; $\alpha = 1, \cdots, M_n$;
$n = 1, \cdots, \infty$; and the centers $\lambda_\alpha^n$ are real.
The total number of $\lambda$ variables is $M = \sum_{n=1}^\infty n M_n$.
This hypothesis is believed to be true for
$H=0^+$. (See, e.g., Ref. \refref{destri/lowenstein}.)
As discussed in the Introduction, we expect that our calculation, which is
analogous to the one of Takhtajan${}^{\refref{takhtajan}}$, will lead to
the unique $S$ matrix.

Forming products of the BA equations over the imaginary parts of the
strings (following Takahashi${}^{\refref{takahashi1}}$ and
Gaudin${}^{\refref{gaudin}}$), and then taking the logarithm,
we obtain the following equations for
the string centers:
$$ h_n ( \lambda_\alpha^n ) = J_\alpha^n \,,
\qquad \alpha = 1, \cdots, M_n\,, \quad n = 1, \cdots, \infty \,,
\eqlabel{BAlog} $$
where the functions $h_n (\lambda)$ are defined by
$$h_n (\lambda) = {1\over 2\pi} \left\{
\left[ N q_n(\lambda) + \Xi_{n 1}(\lambda) \right]
-\sum_{m=1}^\infty \sum_{\beta=1}^{M_m}  \Xi_{n m}
(\lambda - \lambda_\beta^m) \right\} \,, \eqlabel{hn} $$
$q_n (\lambda)$ are odd monotonic-increasing functions defined by
$$q_n (\lambda) = \pi
+ i\log \left( {\lambda + {in\over 2}\over \lambda - {in\over 2}}\right)
\,, \qquad -\pi < q_n (\lambda) \le \pi \,, \eqnum $$
and $\Xi_{n m} (\lambda)$ are given by
$$ \Xi_{n m} (\lambda) = (1 - \delta_{n m}) q_{|n-m|}(\lambda)
+ 2q_{|n-m|+2}(\lambda)
+ \cdots + 2q_{n+m-2}(\lambda) + q_{n+m}(\lambda) \,. \eqnum $$
Moreover, $J_\alpha^n$ are integers or half-odd integers which satisfy
$$|J_\alpha^n| \le J_{max}^n =
N \left( {3\over 2} - {1\over 2}\delta_{n 1} \right)
+ {1\over 2}\left( M_n - 1 \right) - \sum_{m=1}^\infty min (m,n)\ M_m
\,. \eqlabel{range} $$
We obtained the values of $J_{max}^n$ using the prescription of
Faddeev and Takhtajan${}^{\refref{faddeev/takhtajan}}$.
We assume that the numbers $\{ J_\alpha^n \}$ can be regarded as quantum
numbers of the model: for every set $\{ J_\alpha^n \}$ in the range
\eqref{range} (no two of which are identical),
there is a unique solution $\{ \lambda_\alpha^n \}$ (no two of which are
identical) of \eq\eqref{BAlog}.

For the ground state, $M_1 = M_2 = N/2$ and $M_n = 0$ for $n>2$.
Evidently, this spin-singlet state consists of a sea of 1-strings
and a sea of 2-strings. The seas are filled (i.e., there are no holes),
and hence are described for $N \rightarrow \infty$ by root densities
$\rho_n(\lambda) = N^{-1} dh_n(\lambda)/d\lambda \,, n=1,2\,,$ which
can be shown to be given by
$$\rho_1(\lambda) = \rho_2 (\lambda) = s(\lambda) \,, \eqnum $$
where $s(\lambda)= (2 \ch \pi \lambda)^{-1}$. (See also Ref. \refref{we}.)

Excited states of $\nu$ kinks consist of $\nu_1$ holes in the
sea of 1-strings, $\nu_2$ holes in the sea of 2-strings, with
$\nu = \nu_1 + \nu_2$, and a finite number of strings of length $n \ge 3$.
The numbers $\nu_1$ and $\nu_2$ are restricted to be
even non-negative integers. The hole rapidities
$\tilde\lambda_\alpha^n$ are given by
$$ h_n ( \tilde\lambda_\alpha^n ) = \tilde J_\alpha^n \,,
\qquad \alpha = 1 \,, \cdots \,, \nu_n \,, \qquad n = 1\,,2 \,,
\eqlabel{holes} $$
where $\tilde J_\alpha^n$ are integers or half-odd integers which satisfy
$| \tilde J_\alpha^n | \le J_{max}^n$ and which are {\it not} in the
set $\{ J_\alpha^n \}$. Correspondingly, we introduce hole densities
$$\tilde\rho_n(\lambda) = {1\over N} \sum_{\alpha=1}^{\nu_n}
\delta(\lambda - \tilde\lambda_\alpha^n) \,, \qquad n = 1\,,2 \,,
\eqlabel{holedensity} $$
and string densities
$$\rho_n(\lambda) = {1\over N} \sum_{\alpha=1}^{M_n}
\delta(\lambda - \lambda_\alpha^n) \,, \qquad n \ge 3 \,.
\eqlabel{stringdensity} $$
The densities $\rho_1$ and $\rho_2$ now satisfy
$$\rho_n(\lambda) + \tilde\rho_n(\lambda)
= {1\over N} {d h_n(\lambda)\over d\lambda} \,, \qquad n=1,2\,,
\eqlabel{constraint0} $$
or equivalently${}^{\refref{we}}$
$$\tilde \rho_n +  \sum_{m=1}^\infty A_{n m} * \rho_m = a_n
+ \sum_{l=1}^{min(n,2)} a_{n+3 -2l} \,, \qquad n = 1\,, 2 \,,
\eqlabel{constraint}$$
where
$$a_n(\lambda) = {1\over 2\pi} {d q_n (\lambda)\over d\lambda}
= {1\over 2\pi} {n\over \lambda^2 + {n^2\over 4}} \,, \eqnum $$
$$A_{n m}(\lambda) = \delta_{n m}\delta(\lambda)
+ {1\over 2\pi} {d \Xi_{n m}(\lambda)\over d\lambda} \,, \eqnum $$
and $*$ denotes the convolution
$ \left( f * g \right) (\lambda) = \int_{-\infty}^\infty
d\lambda'\ f(\lambda - \lambda') g(\lambda') \,.$
The linear integral equations \eqref{constraint} can be solved with the
help of Fourier transforms. The solution is
$$\eqalignno{
\rho_1  &= s - \tilde \rho_1 + s * \tilde \rho_2  \,, \cr
\rho_2  &= s + s * \tilde \rho_1 + \sigma * \tilde \rho_2 - \sum_{m=3}^\infty
a_{m-2} * \rho_m \,, \eqalignlabel{solution} \cr} $$
where
$$ \sigma (\lambda) = - {1\over 2\pi} \int_{-\infty}^\infty
{e^{-i \omega \lambda} \over 1 + e^{-|\omega|}} d\omega
\,, \eqnum $$
and the hole ($\tilde\rho_n$) and string ($\rho_n\,, n\ge 3$) densities
are given by Eqs. \eqref{holedensity} and \eqref{stringdensity}, respectively.
The string positions $\lambda_\alpha^n\,, n\ge 3\,, $ are still to be
determined
in terms of the hole positions
$\tilde\lambda_\alpha^1\,, \tilde\lambda_\alpha^2 \,, $
using \eq\eqref{BAlog} for $n \ge 3$.

The energy and momentum of the excitations follows from Eqs.
\eqref{energy} and \eqref{momentum}, respectively:
$$\eqalignno{E &= N\left\{ {7\over 12}
- \pi\sum_{n=1}^\infty \int_{-\infty}^\infty
\left[ a_n + \sum_{l=1}^{min(n,2)} a_{n+3-2l} \right] \rho_n \ d\lambda
\right\} \cr
&= E_0 + \sum_{n=1}^2 \sum_{\alpha =1}^{\nu_n}
\varepsilon(\tilde \lambda_\alpha^n) \,, \eqalignnum \cr}$$
and
$$\eqalignno{P &= -{N\over 2}\sum_{n=1}^\infty \int_{-\infty}^\infty
\left[ (q_n - \pi)
+ \sum_{l=1}^{min(n,2)} (q_{n+3-2l} - \pi) \right] \rho_n \ d\lambda \cr
&= P_0 + \sum_{n=1}^2 \sum_{\alpha =1}^{\nu_n} p(\tilde \lambda_\alpha^n)
\,, \eqalignnum \cr}$$
where $E_0$ and $P_0$ are the energy and momentum of the ground state,
respectively, and $\varepsilon(\lambda)$ and $p(\lambda)$ are the
energy and momentum of a kink with rapidity $\lambda$,
$$\varepsilon(\lambda) = \pi s(\lambda) = {\pi \over 2 \ch \pi \lambda}
\,, \qquad
p(\lambda) = -{\pi\over 4} + {1\over 2} \tan^{-1} \sh \pi \lambda
\,. \eqlabel{kinkem} $$
We observe that
$${d p \over d\lambda} = \varepsilon (\lambda)\,, \eqlabel{observe} $$
and that the kink dispersion relation is
$$\varepsilon = -{\pi\over 2} \sin 2p \,, \eqnum $$
implying that the (unique) speed of sound is $v_s = \pi$.

A useful formula for the spin is
$$S^z = {\nu_2\over 2} - \sum_{n=3}^\infty (n-2) M_n \,,
\eqlabel{nice} $$
which follows from \eqref{spin}, and from the fact that
$$M_1 ={N\over 2} - \nu_1 + {\nu_2\over 2} \,, \qquad
  M_2 ={N\over 2} + {\nu_1 - \nu_2\over 2} - \sum_{n=3}^\infty M_n \,.
\eqnum $$
Notice that the expression \eqref{nice} for $S^z$ is independent of $\nu_1$,
which implies that the kinks corresponding to holes in the sea of 1-strings
have spin 0. We shall discuss this point further below.

We particularize now to the case of two-kink excitations $(\nu = 2$).
There are three possibilities:

\noindent
(i) {\it triplet} $(S^z =1)$

This state is characterized by $M_1 ={N\over 2} + 1$; $M_2 ={N\over 2} - 1$;
$M_n = 0$ for $n>2$, which corresponds to two holes in the sea of 2-strings
($\nu_1 = 0 \,, \nu_2 = 2$). The root densities are given by
(see \eq\eqref{solution})
$$\eqalignno{
\rho_1 (\lambda) &= s(\lambda) + {1\over N}\sum_{\alpha=1}^2
s(\lambda - \tilde\lambda_\alpha^2) \,, \cr
\rho_2 (\lambda) &= s(\lambda) + {1\over N}\sum_{\alpha=1}^2
\sigma(\lambda - \tilde\lambda_\alpha^2) \,. \eqalignlabel{triplet}
\cr} $$

\medskip

\noindent
(ii) {\it singlet} $(S^z =0)$

This state is characterized by $M_1 ={N\over 2} + 1$; $M_2 ={N\over 2} - 2$;
$M_3 = 1$; $M_n = 0$ for $n>3$, which corresponds to two holes in the sea
of 2-strings ($\nu_1 = 0 \,, \nu_2 = 2$) and one 3-string. The root densities
are given by
$$\eqalignno{
\rho_1 (\lambda) &= s(\lambda) + {1\over N}\sum_{\alpha=1}^2
s(\lambda - \tilde\lambda_\alpha^2) \,, \cr
\rho_2 (\lambda) &= s(\lambda) + {1\over N}\left[ \sum_{\alpha=1}^2
\sigma(\lambda - \tilde\lambda_\alpha^2) - a_1(\lambda - \lambda_1^3) \right]
\,, \eqalignlabel{singlet1}
\cr} $$
where $\lambda_1^3 = (\tilde\lambda_1^2 + \tilde\lambda_2^2)/2$.

\medskip

\noindent
(iii) {\it singlet} $(S^z =0)$

This state is characterized by $M_1 ={N\over 2} - 2$; $M_2 ={N\over 2} + 1$;
$M_n = 0$ for $n>2$, which corresponds to two holes in the sea of 1-strings
($\nu_1 = 2 \,, \nu_2 = 0$). The root densities are given by
$$\eqalignno{
\rho_1 (\lambda) &= s(\lambda) - {1\over N}\sum_{\alpha=1}^2
\delta(\lambda - \tilde\lambda_\alpha^1) \,, \cr
\rho_2 (\lambda) &= s(\lambda) + {1\over N}\sum_{\alpha=1}^2
s(\lambda - \tilde\lambda_\alpha^1) \,. \eqalignlabel{singlet2}
\cr} $$

All together, there are 5 two-kink states.
Taking into account degeneracies from the possible values of the string
positions, we find that the number of $\nu$-kink states is given by
$$\sum_{m=0}^{\nu/2} 2^{2m}    \,. \eqlabel{count} $$
In comparison, for the spin 1/2 Heisenberg chain, there are only 4
two-kink states (triplet plus singlet); and the number of $\nu$-kink
states is only $2^\nu$.

Our interpretation of these results is that there are two types of kinks:
the usual spin 1/2 kink and a new spin 0 kink. We refer to these as
spinor kinks and scalar kinks, respectively. The number of scalar kinks
is $\nu_1$, and the number of spinor kinks is $\nu_2$.
As already remarked, multi-kink states are subject to the super-selection rule
that both $\nu_1$ and $\nu_2$ must be even.
Evidently, configurations of kinks which differ only by interchanges
of scalars with spinors are not distinct. Hence, for given values of $\nu_1$
and $\nu_2$, the number of states is $2^{\nu_2}$.
With the above assignment of spin quantum numbers to the kinks, and with the
super-selection rule, we reproduce the counting \eqref{count} of $\nu$-kink
states. There is no evidence of RSOS structure.

\bigskip

\noindent
{\bf 3. $S$ matrix}

\bigskip

We compute the $S$ matrix following the method of
Korepin${}^{\refref{korepin}}$ in a formulation given by
Andrei and Destri${}^{\refref{andrei/destri}}$. The basic
idea can be understood by considering a particle moving in one dimension
with momentum $p$ in an even finite-range potential. The $S$ matrix is diagonal
in the basis of parity eigenstates, with matrix elements $S = e^{i \phi}$.
Putting the system in a periodic box of length $L$ leads to the quantization
condition
$$e^{i p L} S = 1 \,. \eqnum $$
Therefore, the momentum of the interacting particle is related to the
phase shift $\phi$ by
$$p = {2\pi\over L} m - {1\over L} \phi \,, \eqlabel{quantization} $$
where $m$ is an integer. With these conventions, the phase shift for an
attractive potential is positive.

The same formula holds for the scattering of kinks, with $L = 2N$ being
the number of spins in the chain. In order to obtain the phase shifts,
we must identify in the expression for the momentum
$p(\tilde\lambda_\alpha^n)$ of a kink with rapidity $\tilde\lambda_\alpha^n$
the free term $2\pi m/L = \pi m/N$. We can
accomplish this with the help of the counting function $h_n (\lambda)$,
a convenient expression for which is obtained
by integrating \eq\eqref{constraint0}
$${1\over N} h_n(\lambda) = {1\over 2}\int_{-\infty}^\infty
\epsilon(\lambda - \lambda')\
\left[ \rho_n(\lambda') + \tilde\rho_n(\lambda') \right]\ d\lambda'
+ c_n \,,\qquad n=1,2\,,
\eqnum $$
where $\epsilon(x)= {\hbox{ sign }} x = x/|x|$. We henceforth drop the
integration constants $c_n$, which contribute only additive constants
to the phase shifts. Evaluating the counting
function at $\tilde\lambda_\alpha^n$ gives the integer or half-odd
integer $\tilde J_\alpha^n$, as we see from \eq\eqref{holes}. Therefore,
$${1\over N}\tilde J_\alpha^n  = {1\over 2}\int_{-\infty}^\infty
\epsilon(\tilde\lambda_\alpha^n  - \lambda)\
s(\lambda) d\lambda
+ {1\over 2}\int_{-\infty}^\infty
\epsilon(\tilde\lambda_\alpha^n - \lambda)\ r_n(\lambda)\ d\lambda
 \,, \qquad n=1,2\,,
\eqlabel{i1} $$
where the quantities $r_n(\lambda)$ are defined by
$$ \rho_n(\lambda) + \tilde\rho_n(\lambda) = s(\lambda) + r_n(\lambda)
\,, \qquad  n=1,2\,. \eqnum $$

There remains to relate $\tilde J_\alpha^n$ to the momentum
$p(\tilde\lambda_\alpha^n)$ of a kink.
{}From Eqs. \eqref{kinkem} and \eqref{observe}, we have that
$$p(\tilde\lambda_\alpha^n) = {\pi\over 2}\int_{-\infty}^\infty
\epsilon(\tilde\lambda_\alpha^n  - \lambda)\
s(\lambda)\ d\lambda  - {\pi\over 4} \,. \eqlabel{i2} $$
We conclude from Eqs. \eqref{quantization}, \eqref{i1}, \eqref{i2}
that the phase shifts are given (up to additive constants) by
$$\phi(\tilde\lambda_\alpha^n) = \pi N \int_{-\infty}^\infty
\epsilon(\tilde\lambda_\alpha^n - \lambda)\ r_n(\lambda)\ d\lambda
\,, \qquad n=1,2 \,. \eqlabel{phaseshift} $$

For two-kink states, this expression is a function of
$\tilde \lambda_{12}^n \equiv \tilde \lambda_1^n - \tilde \lambda_2^n$,
which we shall take to be positive. In order to evaluate the above
integral, we follow Faddeev and Takhtajan, and first compute its derivative:
$${d\phi(\tilde \lambda_{12}^n)\over d \tilde \lambda_{12}^n}
= 2 \pi N r_n (\tilde\lambda_1^n) \,. \eqlabel{derivative} $$
This procedure introduces further integration constants in the expression
for the phase shift, which could in principle be determined
by directly evaluating \eqref{phaseshift}. The evaluation of
\eqref{derivative} is accomplished with the help of Fourier transforms
of $s(\lambda)$ and $a_n(\lambda)$, the only nontrivial integral which
appears being
$$\int_0^\infty {\cos \lambda \omega \over e^\omega + 1} d\omega
= {1\over 4}\left[ \psi({i\lambda\over 2}) + \psi(-{i\lambda\over 2})
- \psi({1\over 2} + {i\lambda\over 2}) - \psi({1\over 2} - {i\lambda\over 2})
\right]   \,, \eqnum $$
where $\psi(z) = d \log \Gamma (z)/dz$.

For the triplet state \eqref{triplet}, we obtain the $S$ matrix
$$S_t(\tilde\lambda_{12}^2) = {\Gamma({i\tilde\lambda_{12}^2\over 2})
\Gamma({1\over 2} - {i\tilde \lambda_{12}^2\over 2})\over
\Gamma(- {i\tilde \lambda_{12}^2\over 2})
\Gamma({1\over 2} + {i\tilde \lambda_{12}^2\over 2})}
\,; \eqlabel{smatrix/triplet}$$
for the first singlet state \eqref{singlet1}, we obtain
$$S_{s1}(\tilde \lambda_{12}^2) = {\tilde \lambda_{12}^2 + i\over
\tilde \lambda_{12}^2 - i} S_t(\tilde \lambda_{12}^2)
\,; \eqlabel{smatrix/singlet1} $$
and for the second singlet state \eqref{singlet2}, we obtain
$$S_{s2}(\tilde \lambda_{12}^1) = 1
\,; \eqlabel{smatrix/singlet2} $$
up to constant (rapidity-independent) phase factors.

The triplet and first singlet $S$ matrices \eqref{smatrix/triplet},
\eqref{smatrix/singlet1} coincide with those for the Heisenberg
chain${}^{\refref{faddeev/takhtajan}}$ \footnote \dag{Actually, the phase
shifts of Faddeev-Takhtajan have the opposite sign. One can recover the
Faddeev-Takhtajan results by invoking the prescription${}^{\refref{korepin}}$
that the phase shift for the scattering of a hole with another hole
(or with a particle) acquires an additional minus sign.
Correspondingly, the physical strip for a hole becomes
$-1 < {\hbox{ Im }} \tilde\lambda \le 0$.}. That is, the spinor
kinks of the alternating chain have the same scattering as the kinks
of the spin 1/2 chain. The second singlet $S$ matrix \eqref{smatrix/singlet2}
is trivial. Evidently, there is no scattering between scalar kinks.

There remains to determine whether there is scattering between
the spinor kinks and the scalar kinks. Since a two-kink state
consisting of a spinor kink and a scalar kink is not possible,
one must consider states with at least four kinks in order to answer
this question. We have analyzed the four-kink state characterized by
$M_1 ={N\over 2} - 1$; $M_2 ={N\over 2}$; and $M_n = 0$ for $n>2$, for
which $\nu_1 = \nu_2 = 2$\footnote*{Factorization implies that
an analysis of other states with $\nu_1\,, \nu_2 \ge 2$ should give
the same result.}. We find that the $S$ matrix for scalar - spinor
scattering is given by
$$S(\tilde\lambda) = i \coth {\pi\over 2}
\left( \tilde\lambda + {i\over 2}\right)
\,, \eqlabel{smatrix/spinor} $$
where $\tilde\lambda = \tilde\lambda_\alpha^1 - \tilde\lambda_\beta^2$.
The pole at $\tilde\lambda = -i/2$ does not lie on the physical sheet
$0 \le {\hbox{ Im }} \tilde\lambda < 1$, and so does not correspond to a
bound state. Evidently, there is nontrivial scattering between a scalar and
a spinor.

The $S$ matrix \eqref{smatrix/triplet} - \eqref{smatrix/spinor}
is crossing-symmetric and unitary.

\bigskip

\noindent
{\bf 4. Discussion}

\bigskip

We have emphasized that the property \eqref{commute} is satisfied for
the alternating spin 1/2 and spin 1 chain given by \eq\eqref{hamiltonian}.
There is no ambiguity (at least for the central charge) at the point
$T = H = 0$. Therefore, it is plausible that results obtained at this
point (even using the string hypothesis) should be unique.
The property \eqref{commute} is also satisfied by the $su(n)$ chain
in the fundamental representation (see Ref. \refref{su}). It would be
interesting to find other models for which this property is satisfied.

We have seen that the chain \eqref{hamiltonian} has gapless kinks
(``spin waves'') which have spin 0, as well as spin 1/2. To the best of
our knowledge, this is the first example of a magnetic chain with scalar
kinks. Remarkably, the scalar-spinor scattering is nontrivial, yet the
spinor-spinor scattering is the same as if the scalar kinks were absent.
Moreover, there is no scalar-scalar scattering.

It would be interesting to determine the conformal field theory (CFT)
to which this model corresponds. Certainly, the CFT must be
$su(2)$-invariant, and have central charge $c=2$. One can rule out
the level-four $SU(2)$ WZW model, since this CFT corresponds to the
homogeneous $s=2$ chain, whose excitations and $S$
matrix${}^{\refref{takhtajan},\refref{reshetikhin}}$
are very different from what we have found here. Another
possibility${}^{\refref{aladim/martins1}, \refref{martins}}$
is a set of two level-one $SU(2)$ WZW models. However, this CFT
also does not seem to be compatible with our set of excitations.

The $S$ matrix \eqref{smatrix/triplet} - \eqref{smatrix/spinor}
can also be obtained by starting with an anisotropic
chain${}^{\refref{devega/woynarovich}}$ with anisotropy parameter
$\eta$ and lattice spacing $a$, and then taking the continuum limit
$a \rightarrow 0$ and the isotropic limit $\eta \rightarrow 0$
while keeping a mass parameter
$m^2 \propto a^{-2} \exp (- \pi^2/\eta)$ fixed. (See Ref. \refref{mccoy/wu}.)
Thus, this $S$ matrix should also be described by
a massive $su(2)$-invariant integrable quantum field theory,
which in the ultraviolet limit $m \rightarrow 0$ reduces to the
above-mentioned CFT.

\bigskip

We thank C.Destri, F. E{\ss}ler, V. Korepin, A. Leclair, and N. Reshetikhin
for valuable discussions.
One of us (H J de V) thanks the Department of Physics, University
of Miami, for its hospitality; and two of us (LM and RN) thank the
LPTHE and University of Bonn for their hospitality. Part of this work was
performed at the Aspen Center for Physics. This work was supported in part
by the National Science Foundation under Grant No. PHY-92 09978.

\bigskip

\noindent
{\bf References}

\vskip 0.2truein

\reflabel{bethe}
H. Bethe, Z. Phys. {\it 71} (1931) 205.

\reflabel{faddeev/takhtajan}
L.D. Faddeev and L.A. Takhtajan, J. Sov. Math. {\it 24}, 241 (1984).

\reflabel{korepin}
V.E. Korepin, Theor. Math. Phys. {\it 76} (1980) 165;
V.E. Korepin, G. Izergin and N.M. Bogoliubov, {\it Quantum Inverse
Scattering Method, Correlation Functions and Algebraic Bethe Ansatz}
(Cambridge University Press, 1993); F.H.L. E{\ss}ler and V.E. Korepin,
``$su(2) \times su(2)$-invariant scattering matrix of the Hubbard model''

\reflabel{takahashi1}
M. Takahashi, Prog. Theor. Phys. {\it 46} (1971) 401.

\reflabel{gaudin}
M. Gaudin, Phys. Rev. Lett. {\it 26} (1971) 1301;
{\it La fonction d'onde de Bethe} (Masson, 1983).

\reflabel{johnson/mccoy}
J.D. Johnson and B.M. McCoy, Phys. Rev. {\it A6}, 1613 (1972).
For a recent review, see L. Mezincescu and R.I. Nepomechie, in
{\it Quantum Groups, Integrable Models and Statistical Systems},
ed. by J. Le Tourneux and L. Vinet (World Scientific), in press.

\reflabel{takahashi2}
M. Takahashi, Prog. Theor. Phys. {\it 50} (1973) 1519;
{\it 51} (1974) 1348.

\reflabel{filyov}
V.M. Filyov, A.M. Tsvelik and P.B. Wiegmann, Phys. Lett. {\it 81A}
(1981) 175.

\reflabel{bpz}
A.A. Belavin, A.M. Polyakov and A.B. Zamolodchikov, Nucl. Phys. {\it B241}
(1984) 333.

\reflabel{critical}
H.W.J. Bl\"ote, J.L. Cardy and M.P. Nightingale, Phys. Rev. Lett. {\it 56}
(1986) 742; I. Affleck, Phys. Rev. Lett. {\it 56} (1986) 746.

\reflabel{wzw}
E. Witten, Commun. Math. Phys. {\it 92} (1984) 455; S.P. Novikov, Usp. Mat.
Nauk {\it 37} (1982) 3.

\reflabel{zamo}
A.B. Zamolodchikov and Al.B. Zamolodchikov, Ann. Phys. {\it 120} (1979) 253;
Nucl. Phys. {\it B379} (1992) 602; P. Fendley, H. Saleur, and Al. B.
Zamolodchikov, Int. J. Mod. Phys. {\it A}, in press; P. Fendley, Phys. Rev.
Lett., in press.

\reflabel{zamolodchikov/fateev}
A.B. Zamolodchikov and V.A. Fateev, Sov. J. Nucl. Phys. {\it 32} (1980) 298.

\reflabel{kulish/sklyanin}
P.P. Kulish and E.K. Sklyanin, {\it Lecture Notes in Physics} {\it 151}
(Springer, 1982) 61;
P.P. Kulish, N.Yu. Reshetikhin and E.K. Sklyanin, Lett. Math. Phys. {\bf 5}
(1981) 393.

\reflabel{babujian}
H.M. Babujian, Nucl. Phys. {\it B215}, 317 (1983).

\reflabel{takhtajan}
L.A. Takhtajan, Phys. Lett. {\it 87A}, 479 (1982).

\reflabel{unpub}
L. Mezincescu and R.I. Nepomechie, unpublished.

\reflabel{alcaraz/martins}
F.C. Alcaraz and M.J. Martins, J. Phys. {\it A21} (1988) 4397.

\reflabel{klumper}
A. Klumper, M.T. Batchelor and P.A. Pearce, J. Phys. {\it A24} (1991) 3111.

\reflabel{andrei/destri}
N. Andrei and C. Destri, Nucl. Phys. {\it B231} (1984) 445.

\reflabel{faddeev/reshetikhin}
L.D. Faddeev and N. Reshetikhin, Ann. Phys. {\it 167} (1985) 227.

\reflabel{reshetikhin}
N. Reshetikhin, J. Phys. {\it A24} (1991) 3299.

\reflabel{destri/lowenstein}
C. Destri and J.H. Lowenstein, Nucl. Phys. {\it B205} (1982) 369;
A.M. Tsvelick and P.B. Wiegmann, Adv. in Phys. {\it 32} (1983) 453.

\reflabel{devega/woynarovich}
H.J. de Vega and F. Woynarovich, J. Phys. {\it A25}, 4499 (1992).

\reflabel{we}
H.J. de Vega, L. Mezincescu and R.I. Nepomechie, ``Thermodynamics
of Integrable Chains with Alternating Spins,'' Phys. Rev. {\it B},
in press.

\reflabel{aladim/martins1}
S.R. Aladim and M.J. Martins, J. Phys. {\it A26} (1993) L529.

\reflabel{aladim/martins2}
S.R. Aladim and M.J. Martins, J. Phys. {\it A26} (1993) 7287.

\reflabel{martins}
M.J. Martins, J. Phys. {\it A26} (1993) 7301.

\reflabel{pfeffer}
D.M. Pfeffer, ``Critical Properties of the Alternating Spin 1/2 - Spin $s$
Chain''

\reflabel{su}
S.V. Pokrovskii and A.M. Tsvelik, Sov. Phys. JETP {\it 66} (1987) 1275;
H.J. de Vega, J. Phys. {\it A20} (1987) 6023;
F.C. Alcaraz and M.J. Martins, J. Phys. {\it A22} (1989) L865;
L. Mezincescu, R.I. Nepomechie, P.K. Townsend, and A.M. Tsvelik,
Nucl. Phys. {\it B406} (1993) 681.

\reflabel{mccoy/wu}
B.M. McCoy and T.T. Wu, Phys. Lett. {\it 87B} (1979) 50;
H.J. de Vega, Int. J. Mod. Phys. {\it A4} (1989) 2371;
H.J. de Vega, L. Mezincescu and R.I. Nepomechie, in preparation.

\vfill\eject
\end